\documentclass[trackchanges]{aastex701}

\usepackage{CJK}
\usepackage{amsmath,amssymb,amsthm,bm}
\usepackage{graphicx}
\usepackage{color}
\usepackage{soul,xcolor}
\usepackage{float}
\usepackage{caption}

\begin{document}
\begin{CJK*}{UTF8}{gbsn}


\title{Radial Evolution of Correlation Anisotropy in PUNCH Observations of the Young Solar Wind}

\author{Zhuoran Gao}
\affiliation{Department of Physics and Astronomy,
University of Delaware, Newark, DE 19716, USA}
\email{veciam@udel.edu}
\author{Yan Yang}
\affiliation{Department of Physics and Astronomy,
University of Delaware, Newark, DE 19716, USA}
\email{yanyang@udel.edu}
\author[orcid=0000-0003-4168-590X]{Francesco Pecora}
\affiliation{Department of Physics and Astronomy,
University of Delaware, Newark, DE 19716, USA}
\email{fpecora@udel.edu}
\author[orcid=0000-0002-7174-6948]{Rohit Chhiber}
\affiliation{Department of Physics and Astronomy,
University of Delaware, Newark, DE 19716, USA}
\affiliation{Heliophysics Science Division, NASA Goddard Space Flight Center, Greenbelt, MD 20771, USA}
\email{rohitc@udel.edu}
\author{Riley Amos}
\affiliation{Southwest Research Institute, Boulder, CO, USA}
\email{ramos@smith.edu}
\author[orcid=0000-0002-3699-3134]{Steven R. Cranmer}
\affiliation{Department of Astrophysical and Planetary Sciences,
Laboratory for Atmospheric and Space Physics,
University of Colorado, Boulder, CO 80309, USA}
\email{steven.cranmer@colorado.edu}
\author[0009-0008-8723-610X]{Jiaming Wang}
\affiliation{Department of Physics and Astronomy,
University of Delaware, Newark, DE 19716, USA}
\email{jmwang@udel.edu}
\author{William H. Matthaeus}
\affiliation{Department of Physics and Astronomy,
University of Delaware, Newark, DE 19716, USA}
\email{whm@udel.edu}
\author{Craig E. DeForest}
\affiliation{Southwest Research Institute, Boulder, CO, USA}
\email{craig.deforest@swri.org}
\author{Sarah E. Gibson}
\affiliation{NSF National Center for Atmospheric Research / High Altitude Observatory, Boulder, CO, USA}
\email{sgibson@ucar.edu}
\author{Nicholeen Viall}
\affiliation{Heliophysics Science Division, NASA Goddard Space Flight Center, Greenbelt, MD 20771, USA}
\email{}
\author{Chris R. Gilly}
\affiliation{NorthWest Research Associates, Boulder, CO, USA}
\email{gilly@nwra.com}
\author{Ronnie Killough}
\affiliation{Southwest Research Institute, San Antonio, TX, USA}
\email{}
\author{Jillian Redfern}
\affiliation{Southwest Research Institute, Boulder, CO, USA}
\email{}
\author{William L. Wells}
\affiliation{Southwest Research Institute, San Antonio, TX, USA}
\email{}
\author{Matthew N. Beasley}
\affiliation{Southwest Research Institute, Boulder, CO, USA}
\email{}
\author{Glenn T. Laurent}
\affiliation{Southwest Research Institute, Boulder, CO, USA}
\email{}
\author{Robin C. Colaninno}
\affiliation{Naval Research Laboratory, Washington, DC, USA}
\email{}
\author{Nick R. Waltham}
\affiliation{United Kingdom Research \& Innovation / Science \& Technology Facilities Council / Rutherford Appleton Laboratory, Harwell Campus, Oxfordshire, U.K.}
\email{}
\author{J. Marcus Hughes}
\affiliation{Southwest Research Institute, Boulder, CO, USA}
\email{}
\author{Daniel B. Seaton}
\affiliation{Southwest Research Institute, Boulder, CO, USA}
\email{}
\author{Samuel Van Kooten}
\affiliation{Southwest Research Institute, Boulder, CO, USA}
\email{}
\author{Ritesh Patel}
\affiliation{Southwest Research Institute, Boulder, CO, USA}
\email{}
\author{Chris Lowder}
\affiliation{Southwest Research Institute, Boulder, CO, USA}
\email{}
\author{Derek Lamb}
\affiliation{Southwest Research Institute, Boulder, CO, USA}
\email{}


\begin{abstract}
Solar wind turbulence, 
evolving in the presence of a mean magnetic field, large-scale shear, and expansion, exhibits several types of anisotropy, including wavevector anisotropy and variance anisotropy. As the solar wind expands from the corona into the interplanetary space, the magnetically controlled outer corona transitions into the dynamic and less constrained solar wind, and fluctuations may evolve toward isotropy. This is suggestive of an evolution of 
anisotropy with distance from the Sun, 
a feature observed in 
earlier results using surveys of in-situ data, as well as more limited images of solar wind density from STEREO.
Here we confirm and expand on those results
employing data from the PUNCH mission, 
whose wide-field, high-cadence imaging enables anisotropy tracking at different heliospheric positions. We evaluate a two-dimensional (2D) correlation function $C(\ell_{\perp}, \ell_{\parallel})$ of polarized white-light brightness as a function of radial $\ell_\parallel$ and transverse $\ell_\perp$ scales to infer the nature of density anisotropy
and its development in space. At larger heliocentric distances (\(\sim 100 R_\odot\)), circular contours of the correlation function indicate a transition toward approximate isotropy. The correlation structure at heliocentric distances from $30R_\odot$ to $60R_\odot$ preferentially extends along the radial direction, which might imply the possible dominance of the two-dimensional variation in the transverse plane. In this way, the very large field of view of PUNCH images reveal a fundamental transition of solar wind turbulence toward isotropy.
\end{abstract}




\section{Introduction}
\label{sec:introduction}

Highly dynamic or turbulent fluids and plasmas
exhibit deviations
from statistical isotropy
(i.e., anisotropy) due to expansion~\citep[e.g.,][]{VolkAlpers73,GrappinVelli96},
and 
gradients of several types
\citep{DeLucaEA90,BiferaleProcaccia05}. 
Anisotropy of the turbulent fluctuation can also be induced in the 
presence of preferred directions
such as those associated with rotation, for example, solar rotation
\citep{MininniPouquet10rotating},
or an imposed mean magnetic field 
\citep{ShebalinEA83}, effects that cannot be removed by transformation of 
the inertial frame \citep{Speziale1989}.
The vast expanse of literature 
on the causes of anisotropy
\citep[e.g., reviewed by ][]{OughtonEA15}
makes it clear that
not only 
numerous physical processes may be responsible, 
but also the nature of anisotropy may vary over different sizes and length scales,
and that the causes
may vary at different scales and in different systems. 
An early report of 
anisotropy at the scale of a plasma experimental 
device \citep{RobinsonRusbridge71}
provided motivation to examine analogous 
effects on correlations in magnetohydrodynamics (MHD) simulation
associated with
the magnetic field direction
\citep{ShebalinEA83,OughtonEA94}.
The first observational evidence of correlation anisotropy in the solar wind \citep{CrookerEA82}
employed 1~AU (astronomical units) magnetic field data from ISEE 1 and ISEE 3. 

Spectral (or correlation) anisotropy is 
often associated with cross-scale energy transfer that is also anisotropic. In particular, for incompressible MHD, there is abundant evidence \citep{OughtonEA15,VerdiniEA15,JiangEtAl2023_JFM_hypervisc} that strong (and even moderate) mean magnetic fields compared to fluctuations significantly suppress energy transfer in the parallel direction with respect to the mean magnetic field, so that the perpendicular energy cascade, being
unaffected by the mean field,
is much stronger than the parallel cascade. Note that spectral anisotropy usually occurs over a range of scales, rather than at isolated, individual scales.
The solar wind can exhibit several types of scale-dependent anisotropy in response to expansion, a mean magnetic field, and large-scale shear.
At the local turbulence scales, on the order of the correlation length or less, 
wavevector anisotropy and variance anisotropy \citep{ShebalinEA83,HorburyEA12,OughtonEA15}
may be associated with local MHD scale processes \citep{Barnes76,ShebalinEA83}.
Meanwhile at larger scales, 
effects of more global origin, such as
expansion, and stream interactions that produce both shear and compression, 
may be the dominant factors in 
generating anisotropy \citep{RobertsEA92-jgr,DeForestEA16,DeForestEA18,WangEA24-density}.

The development of anisotropy in solar wind fluctuations has been observed using surveys of in situ data (e.g., \citep{CuestaEA22-iso,bandyopadhyay2021geometry}) as well as limited remote sensing observations using STEREO \citep{DeForestEA16}.
The PUNCH mission affords a unique opportunity 
to observe structures and emerging properties of the heliospheric plasma through 
high resolution 
polarized white-light imaging \citep{deforest2026polarimeter},
thus extending previous preliminary 
correlation analyses of STEREO images 
\citep{DeForestEA16}.

It is important to recognize at the onset that analysis of white-light images differs substantially in its strengths and limitations relative to more familiar in situ plasma and magnetic field observations. The latter is necessarily {\it local}, providing detailed and precise characterization of plasma properties, but limited by sparse 
spatial sampling, even when multiple spacecraft are available. White light studies
on the other hand sample large swaths of the plasma but nonlocal sampling in the direction perpendicular to the plane of the sky 
limits (or at least makes more difficult, see
\citet{PecoraEA26PUNCH}) the interpretation of results. 
The scales covered by PUNCH images range from several {AU}s
to gigameter scales or less, touching on those sensitive to local 
turbulence processes. Note that the correlation scale,
nominally demarcating the large scale end of the inertial range is around 10$^{6}$km at $\sim 1 \rm{AU}$ from the Sun \citep{MatthaeusEA05}.
The present paper has the goal of exploring anisotropy in PUNCH images, making use of PUNCH's unprecedented combination of coverage and resolution.
At this 
preliminary stage
we make only a
cursory attempt to distinguish among the several possible causes of observed scale-dependent
anisotropy.
Specifically, we explore the development of correlation anisotropy in radial and transverse-to-radial directions using several methodological variations and 
correlation functions.
By providing a first view of the development of anisotropy in PUNCH
data, we set the stage for more detailed future 
observational examination of the underlying causes of anisotropy in the interplanetary 
medium.

\section{Strategy and PUNCH datasets}
\label{sec:PUNCH}
 
To quantify the level of anisotropy, studies typically compare the two-point correlation function or its Fourier transform (the spectrum) along different directions, as both quantities measure the energy distribution over scales~\citep{DassoEA05}.
For example, when the energy at a given scale $\ell$ is distributed unevenly with directions, that is defined as spectral (also known as wavevector or correlation) anisotropy. In terms of an energy spectrum $E(k_x,k_y,k_z)$, where $\boldsymbol{k}=(k_x,k_y,k_z)$ is the Fourier wavevector, spectral anisotropy can be represented, for example, by the form $E(k,0,0)\neq E(0,0,k)$, where $k\sim 1/\ell$. Equivalently, one can also consider a homogeneous correlation like $R(\boldsymbol{r}) = \langle \boldsymbol{b}(\boldsymbol{x}) \cdot \boldsymbol{b}(\boldsymbol{x + r}) \rangle$, where $\boldsymbol{b}$ is a fluctuation field, $\boldsymbol{x}$ is the spatial location, and $\boldsymbol{r}$ is the spatial separation. The existence of correlation anisotropy means that $R(r\boldsymbol{\hat{e}}_1) \neq R(r\boldsymbol{\hat{e}}_2)$, for some independent directions $\boldsymbol{\hat{e}}_1, \boldsymbol{\hat{e}}_2$. 
 These fundamental measures of spectral distributions are readily accessible
 using {\it in situ} probes,
 as is well documented in 
 hydrodynamic turbulence literature
 \citep{BatchelorTHT} and in analysis of 
 space mission datasets \citep{Coleman66,MattGold82-rugged,MattEA90}. As mentioned above, such measurements may be very accurate, but their spatial coverage is limited 
 by the location of the probes and turbulence analyses such as the correlation function here depend on the accuracy of approximations such as the Taylor hypothesis.
 Therefore  information about anisotropy may be limited, especially since only a relatively small number of 
 in situ probes is available at any given time.
  Remote sensing
greatly extends the possibilities for investigations, but presents 
 the complication of 
 de-convolving 
 integration effects along 
 limited lines-of-sight.

\begin{figure}[ht]
  \centering
  \includegraphics[width=1\textwidth]{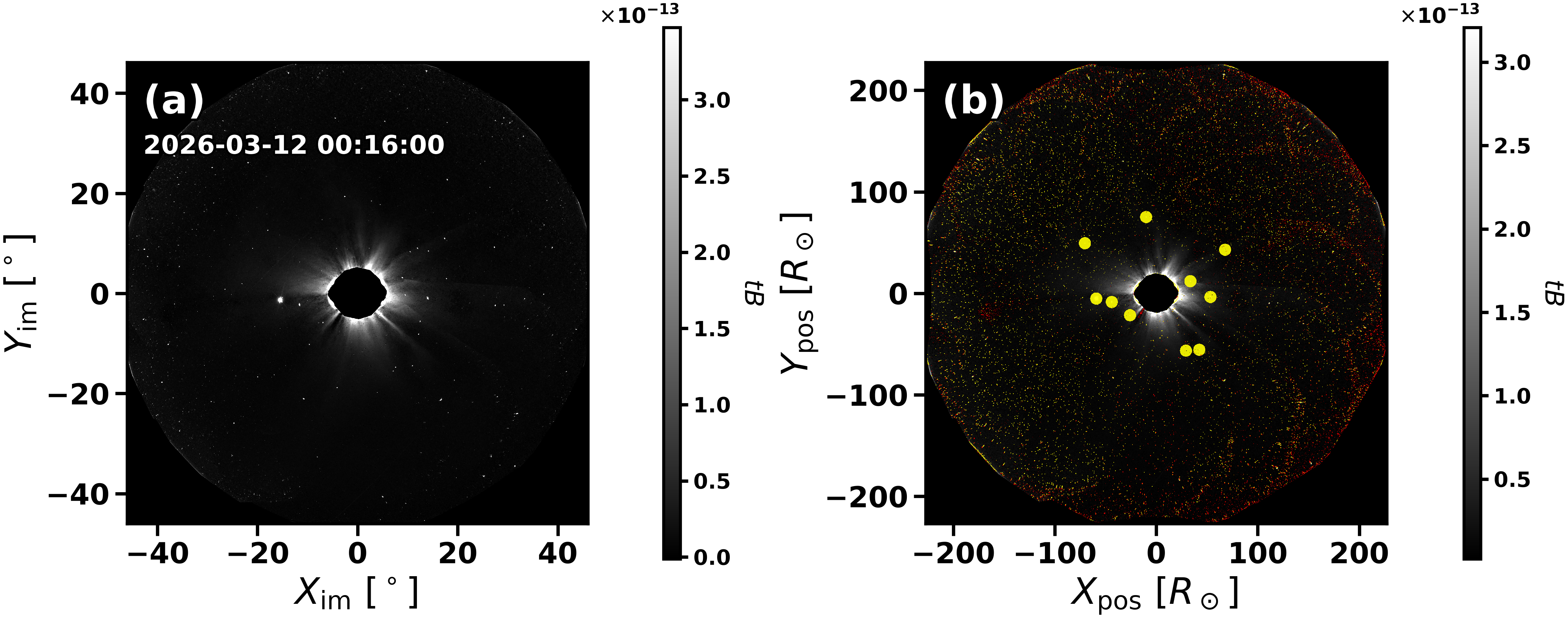}
  \caption{(a) PUNCH total brightness at 00:16:00 UTC on March 12, 2026. (b) Remapped CAM image in the plane of the sky. The yellow-colored and red-colored pixels show the bright and negative-value spots, respectively. Yellow circles are extremely bright features.}
  \label{fig:overview}
\end{figure}

The PUNCH mission \citep{deforest2026polarimeter}
consists of four small spacecraft in Sun-synchronous low Earth orbit, outfitted 
to collect linearly polarized images of the K corona and the young solar wind. One spacecraft carries a Narrow Field Imager (NFI) solar coronagraph \citep{colaninno2025narrow} that captures images of the outer corona at solar elongations from $6R_\odot$ to $32R_\odot$; and the remaining three spacecraft each have a Wide Field Imager (WFI) \citep{laurent2026wide} so that three WFIs together capture a trefoil instantaneous field of view of the sky, at solar elongations from $12R_\odot$ to $180R_\odot$ from the Sun. PUNCH Level 3 data are highly processed and comprise scientific image files. 
Image products are named with a leading ``P'' (Polarized) or ``C'' (Clear (no polarizer)), and are produced at the full observing cadence for that type of image: 4 minutes for P types and 8 minutes for C types.
In this paper, we employ only the WFI dataset, using the 
CAM (Clear 
Averaged 
Mosaic) data product, which provides the field of view by averaging over 32 minutes,
for illustration and analysis. 
Figure \ref{fig:overview}(a) illustrates the total brightness (tB) from CAM data version v0l at 00:16:00 on March 12, 2026, which has the background F-corona removed and the starfield suppressed. 


\subsection{Data Preprocessing}  
Preprocessing of the PUNCH CAM image is required prior to further analysis for the correlation function performed in this work. Several steps, including remapping, bright spots and negative-value cleaning,  are involved in the preprocessing.

PUNCH mosaic images are delivered as azimuthal equidistant projection (AZP) of the celestial sphere, such that the Sun is at the coordinate origin (at the center of the pixel grid). ``Azimuth'' on the image, measured counterclockwise from solar north, corresponds to solar position angle on the sky. 
For the analysis here, we adopt the familiar radial-tangential-normal (RTN) Cartesian system, in which the three axes are: one extending from the Sun to the observer; one extending in the direction of the cross product of solar north with the former axis; and one extending in the mutually perpendicular direction. So the PUNCH image is projected onto the ``plane of the sky'', also called the ``tangent plane'', through $r_{\rm pos} = r_{\rm obs}\tan\epsilon$, where $r_{\rm pos}$ is the projected distance from the Sun, $\epsilon$ is the elongation angle from the Sun and $r_{\rm obs}$ is the Sun-observer distance. More details about PUNCH geometric and photometric quantities can be found in \cite{deforest2026polarimeter}. The 2D image in the plane of the sky, with the resulting Cartesian coordinates $(X_{\rm pos}, Y_{\rm pos})$, is shown in Fig.~\ref{fig:overview}(b). 


There are bright point-source and negative pixels in the data, which may arise from imperfect starfield and background subtraction, or instrumental noise. Negative-value regions, which are marked in red in Fig.~\ref{fig:overview}(b), are excluded during remapping. Bright-source contamination is identified using a two-stage procedure. First, a high-pass filtered image is constructed by subtracting a mask-aware Gaussian background, with the Gaussian weight decreasing with the distance $d$ from each pixel as $w(d)\propto\exp[-d^2/(2\sigma_{\rm HP}^2)]$, where $\sigma_{\rm HP}=0.336R_\odot$. For each pixel in the high-pass filtered image, we evaluate a local median $q_{\rm local}$ and a MAD-based local fluctuation level from $q_{\rm HP}$ over a characteristic spatial scale of approximately $10R_\odot$, and define the parameter
$Z=(q_{\rm HP}-q_{\rm local})/\sigma_{\rm local}$ and
$\sigma_{\rm local}=1.4826\,{\rm MAD}$. The MAD provides a robust
measure of the local fluctuation level that is less sensitive to bright
outliers than the conventional standard deviation, while the factor
1.4826 converts the MAD to the equivalent standard deviation for a
Gaussian distribution. It is trivial that pixels with larger $Z$ will be brighter. 
Pixels with $Z\ge10$, and surrounding pixels with $Z\ge5$ are identified as compact point-source residuals, which are marked as small yellow spots in Fig.\ref{fig:overview}(b).

To identify exceptionally bright sources, we additionally examine a residual image obtained by subtracting a Gaussian-smoothed background.  We adopt the Gaussian weighting form defined above,
with a larger width of $\sigma_{\rm G}=4\,R_\odot$. Between $20R_\odot$ and $100R_\odot$, the image is divided into $5R_\odot$ radial rings. For each ring, we identify the strongest residual source and calculate the fraction of the residual energy within a $2R_\odot$ circle around the source to the total residual energy in the ring. Sources for which this fraction exceeds $30\%$ are classified as exceptionally bright sources and masked with a $6R_\odot$ circle, highlighted as large yellow circles in Fig.\ref{fig:overview}(b). The final bright-source mask is the collection of these regions and the compact-source mask.  These numerical parameters were selected empirically through sensitivity tests, and variations of these parameters do not significantly impact the results here. 
The same parameter set was also tested on images from different observation timestamps and was found to perform consistently across the datasets. 
All excluded pixels, including negative-value and bright-source regions in Fig.~\ref{fig:overview}(b), are removed from subsequent analyses.


\subsection{Background Removal and Radial Falloff}
\begin{figure}[ht]
  \centering
  \includegraphics[width=1\textwidth]{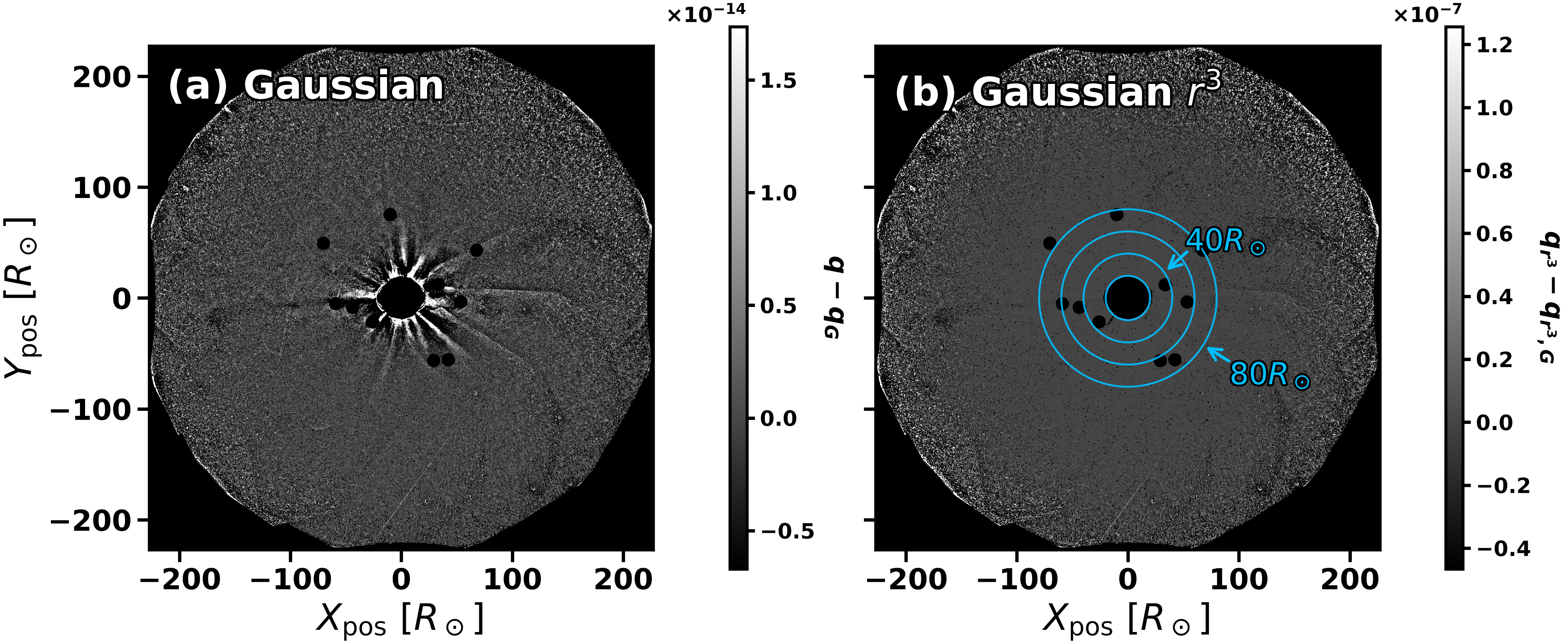}
  \caption{Background-subtracted CAM images at 00:16:00 UTC on March 12, 2026. (a) Residual brightness obtained using Gaussian background subtraction. (b) Residual $r^3$-scaled  brightness. Black regions indicate pixels excluded by the combined bright-source mask. The cyan circles mark heliocentric distances of $20$, $40$, $60$,and $80R_\odot$, indicating the radial annuli used in the correlation analysis.}
  \label{fig:background}
\end{figure}

Before proceeding with the computation of the correlation function, we need to remove the average large-scale radial trend in brightness. A constant or, more generally, a nonzero large-scale background will cause the correlation function to asymptote at a value higher than zero.
After the preprocessing, we remove large-scale background by Gaussian background subtraction. The base image before background removal is denoted by $q$. We construct a local Gaussian background
$q_{\rm G}$ using the Gaussian weighting defined above ($w(d)\propto\exp\left[-d^2/(2\sigma_{\rm G}^2)\right]$), with
$\sigma_{\rm G}=4R_\odot$. The residual field is then obtained by
subtracting the corresponding Gaussian background from the base image. The residual image after background removal is shown in Fig.~\ref{fig:background} (a), 

The base image we are using here for the correlation analysis is a radially scaled version, defined as
$q_{r^3} = r_{\rm pos}^3 q$.
The observed brightness in PUNCH images includes the radial trend of electron density with heliocentric distance and that induced by Thomson scattering \citep{pecora2024magnetohydrodynamic}. 
The density field has a radial falloff proportional to the inverse of the square distance from the Sun \citep{gazis1983radial,hellinger2011heating,perrone2019radial,maruca2023trans,cuesta2023compressible}. 
In addition, Thomson scattering will introduce radial gradients in brightness \citep{billings2013guide}. So the radially scaled brightness is compensated by the radial scaling $r_{\rm pos}^3$.
The radially scaled image after background removal are shown in Fig.~\ref{fig:background}, together with the radial annuli used for the correlation analysis. We compute the correlation function separately within six overlapping radial ranges $[r_{\min}, r_{\max}]$:
$[20\!-\!40],\quad [30\!-\!50],\quad [40\!-\!60],\quad
[50\!-\!70],\quad [60\!-\!80],\quad [70\!-\!90]~R_\odot$.

\section{Correlation Functions}
\label{sec:results}


The normalized correlation function of the radially scaled brightness is computed in each annulus as
\begin{equation*}
    C(\ell_\parallel,\ell_\perp)
=
\frac{
\left\langle q_{r^3}(\mathbf{x})q_{r^3}(\mathbf{x}+\boldsymbol{\ell}) \right\rangle_{\rm annulus}
}{
\frac{1}{4}\sum_{\boldsymbol{\ell}_{\rm ref}}
\left\langle q_{r^3}(\mathbf{x})q_{r^3}(\mathbf{x}+\boldsymbol{\ell}_{\rm ref}) \right\rangle_{\rm annulus}
},
\end{equation*}
where, $\mathbf{x}$ is the pixel position in the
2D image plane, and
$\boldsymbol{\ell}
=
\ell_\parallel\hat{\mathbf{e}}_{\parallel}
+
\ell_\perp \hat{\mathbf{e}}_\perp$
is the lag vector, with $\hat{\mathbf{e}}_{\parallel}$ and
$\hat{\mathbf{e}}_\perp$ denoting the radial and transverse-to-radial unit vectors,
respectively. The average $\langle \cdot \rangle_{\rm annulus}$ is taken
over all valid pixel pairs within each annulus.
Note that the correlation function is normalized with the correlation over a lag $\boldsymbol{\ell}_{\rm ref}$, i.e., $\left\langle q_{r^3}(\mathbf{x})q_{r^3}(\mathbf{x}+\boldsymbol{\ell}_{\rm ref}) \right\rangle_{\rm annulus}$.  A finite reference lag ${\ell}_{\rm ref}$ is used to reduce the influence of residual point-source background, which has significant correlation at very small separations. Using a small finite reference lag therefore reduces their contribution to the normalization.
Here, four reference lag
vectors $\boldsymbol{\ell}_{\rm ref}=(\pm0.5R_\odot,0)$ and
$(0,\pm0.5R_\odot)$ are used.
To compute the correlation function for a given lag vector $\boldsymbol{\ell}$, we need to make use of the fields $q_{r^3}(\mathbf{x})$ and $q_{r^3}(\mathbf{x}+\boldsymbol{\ell})$.  Both fields are required to remain within the same annulus, i.e., $r_{\min} \leq |\mathbf{x}| < r_{\max}$ and $r_{\min} \leq |\mathbf{x}+\boldsymbol{\ell}| < r_{\max}$. The field at
$\mathbf{x}+\boldsymbol{\ell}$ is obtained by interpolation.  

\begin{figure}[ht]
  \centering
  \includegraphics[width=1\textwidth]{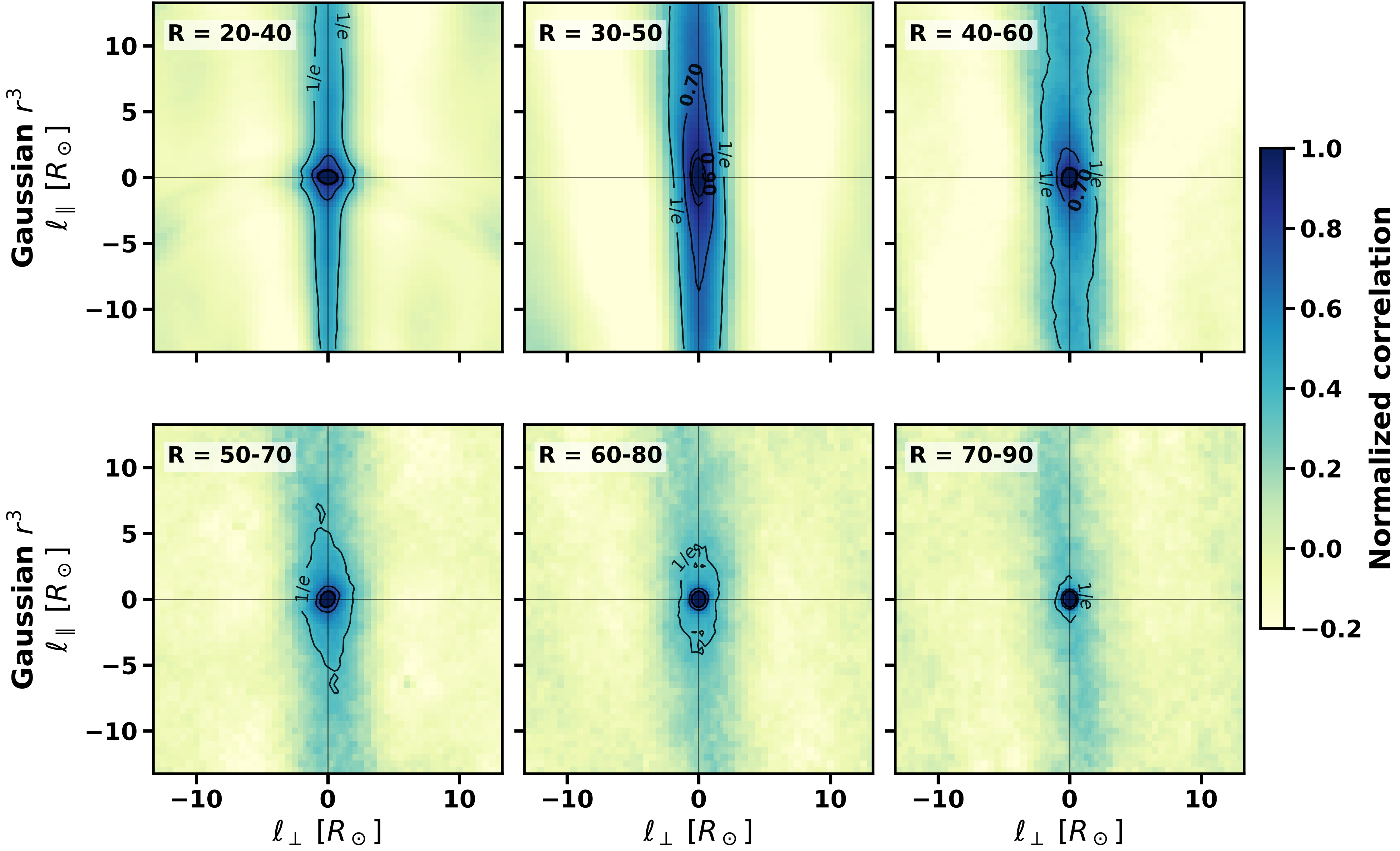}
  \caption{Normalized two-dimensional correlation functions for the radially scaled ($r^3$) brightness in six overlapping radial annuli from $20$ to $90\,R_\odot$. Contours indicate correlation levels of $1/e$, 0.70, 0.90, and 0.95.}
  \label{fig:CAM_overview}
\end{figure}
Figure~\ref{fig:CAM_overview} shows the normalized two-dimensional
correlation functions for the radially scaled brightness. 
In the $20$--$60\,R_\odot$ ranges the correlation functions are elongated
along the radial direction. At larger heliocentric distances, the
radial elongation decreases and the correlation function becomes
more compact. Beyond approximately $60\,R_\odot$, the contours become
more circular, particularly
in the $70$--$90\,R_\odot$ range.
Also noteworthy is the enhanced correlation along the radial direction in the $30$--$50\,R_\odot$ range.

\begin{figure}[ht]
  \centering
  \includegraphics[width=1\textwidth]{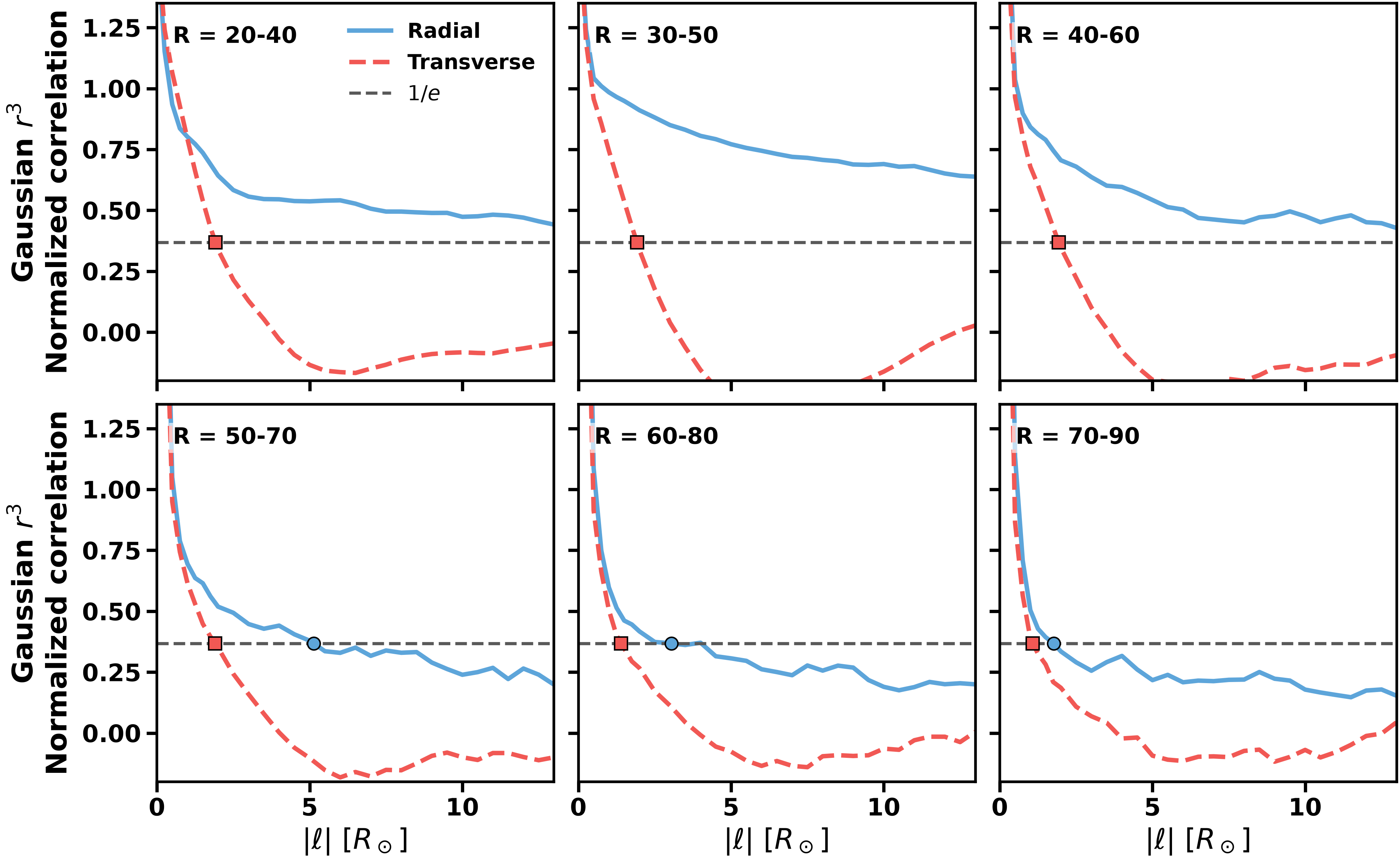}
  \caption{One-dimensional radial and transverse-to-radial correlation profiles for radially scaled ($r^3$) brightness in the six radial ranges. The horizontal dashed line marks the $1/e$ correlation level, and the symbols indicate the corresponding $1/e$ crossing locations.}
  \label{fig:corr_profiles}
\end{figure}

One can also interpret the 2D correlation function in terms of correlation length $\lambda_c$, which can be estimated as the scale over which the correlation function falls down to $1/e$.
The one-dimensional correlation profiles (radial and transverse-to-radial cuts) in
Fig.~\ref{fig:corr_profiles} are derived from the contours in Fig.~\ref{fig:CAM_overview}. We observe several features consistent with Fig.~\ref{fig:CAM_overview}: 
in the $20$--$60\,R_\odot$ ranges, the radial correlation decreases more slowly than transverse-to-radial, and
the larger radial correlation
length is consistent with the radial elongation in Fig.\ref{fig:CAM_overview}.
At larger heliocentric distances, both
correlation lengths decrease and the difference between the two directions
becomes smaller. Again, in the $30$--$50\,R_\odot$ range, a sudden increase of the radial correlation length is observed. 





\section{Discussion and Conclusion}
\label{sec:Discussion}

Data from 
the PUNCH mission
provides a unique 
opportunity to examine
images of electron density
over the 
inner heliosphere, with 
unprecedented resolution
and field of view, which enables us to revisit correlation anisotropy in
heliospheric plasmas here.  
The present analysis reveals a systematic radial evolution of the density correlation function in the inner heliosphere.
The characteristic correlation lengths first increase with heliocentric distance until $30$--$50\,R_\odot$, and then decrease farther from the Sun. At the same time, the difference between the radial and transverse-to-radial correlation lengths becomes progressively smaller, indicating that the correlation structure evolves toward a nearly
isotropic state with increasing heliocentric distance.

Our results show a trend toward isotropy and  $\lambda_c^{\perp}\approx\lambda_c^{\parallel}$ with increasing heliocentric distance below $\approx 90\,R_\odot$, which is consistent with the striated-to-flocculated picture proposed by \cite{DeForestEA16}. \cite{DeForestEA16} analyzed image sequences recorded by the inner Heliospheric Imager instrument on board
the STEREO mission in December 2008, covering distances from approximately $4^\circ$ ($\approx 16 R_\odot$) to $24^\circ$ ($\approx 96 R_\odot$) from the center of the Sun, and identified anomalous fading of the radial striae that
characterize the corona, along with relative brightening of locally-dense puffs of solar wind,
which were described as more isotropic ``flocculae''. This transition has been interpreted as the onset of
dynamical activity associated with velocity shear present in the nascent solar wind plasma
coming from near-radial corotating flux tubes in the corona \citep[][]{RuffoloEA20}. Based on the striation-flocculation transition, the striated structures have long parallel correlation scales, which could also be used to interpret the radial elongated correlation function within $60\,R_\odot$ in Figs.~\ref{fig:CAM_overview} and \ref{fig:corr_profiles}.

As one moves inward closer to the Sun,
our results show $\lambda_c^{\parallel}>\lambda_c^{\perp}$. The trend toward  $\lambda_c^{\parallel}/\lambda_c^{\perp}>1$ at 1~\rm{AU} and beyond has also been measured by in-situ observations \citep[e.g.,][]{SmithEA06-aniso,Podesta09-SWaniso,RuizEA12,CuestaEA22-iso}. However, \cite{CuestaEA22-iso,bandyopadhyay2021geometry} using Parker Solar Probe data found, below $\approx 0.4\rm{AU} (\approx 86 R_\odot)$, a trend toward increasing anisotropy with $\lambda_c^{\parallel}<\lambda_c^{\perp}$ when getting closer to the Sun. Therefore, our result based on PUNCH images is in contrast with PSP observations. This reminds us the slab+2D decomposition \citep[][]{ZankMatt92-sw,ZankMatt93,BieberEA96,MattEA90,ZankEA17}, which represents solar-wind anisotropy as a superposition of slab and two-dimensional (2D) fluctuations. In the slab component, the wavevectors $\boldsymbol{k}$ are assumed to be parallel to the mean magnetic field $\boldsymbol{B}_0$, whereas in the 2D component, $\boldsymbol{k}$ is assumed to be perpendicular to $\boldsymbol{B}_0$. The slab+2D model, though highly idealized, has been remarkably successful in explaining observations. For example, in the sub-Alfv\'enic solar corona, below the Alfv\'en zone, the plasma is essentially quasi-2D \citep{MattEA99-ch,DmitrukEA02,ZankEA18-corona,CranmerWinebarger19}. However, it requires caution to claim that our results of $\lambda_c^{\parallel}>\lambda_c^{\perp}$ support the dominance of the 2D component. The slab+2D picture was defined with respect to the mean magnetic field and verified with in situ measurements of the magnetic field, which contrast with the brightness of the PUNCH image here. Meanwhile, the mean magnetic field (that defines the slab+2D picture) is not radial (parallel direction we are using here) above $20\,R_\odot$ solar radii, and instead tilts 30-45 degrees away from radial at $100\,R_\odot$ \citep{CuestaEA22-iso}. 

Another point that is inferred from our results is the radial evolution of the correlation scale, that is, both the radial $\lambda_c^{\parallel}$ and transverse $\lambda_c^{\perp}$ correlation scales, in general,
decrease with increasing heliocentric distance $R$, except the radial $\lambda_c^{\parallel}$ correlation scale in the $30$--$50\,R_\odot$ range. This appears superficially to be in contrast to the reported results that the correlation scale generally grows with heliocentric distance, from in-situ observations and theoretical models \citep[e.g.,][]{Hollweg86,CranmerEA12,Usmanov2025ApJ}. For example, in-situ statistical surveys of the inner heliosphere suggest that the increase in $\lambda_c$ with $R$ obeys radial power-law fits, $\lambda_c \propto R^\alpha$, with $\alpha\approx0.67$ between $1$ and $10~\rm{AU}$ \citep{matthaeus1999correlation}, $\alpha\approx0.43$ between $0.3$ and $5~\rm{AU}$ \citep{RuizEA14}, $\alpha\approx1.0$ below $0.3~\rm{AU}$, and $\alpha\approx0.3$ beyond \citep{CuestaEA22-iso}. A companion paper \citep{PecoraEA26PUNCH} using PUNCH data estimated the correlation scale as the spectral-break scale separating the energy-containing range from the inertial range in the power spectral density, which also supports an increase in the correlation scale with heliocentric distance. However, we note that the power spectral density in \cite{PecoraEA26PUNCH} corresponds to the azimuthal direction without radial averaging over the $21-50 R_\odot$ range (i.e., the first two radial annuli here). Therefore, the correlation scale reported by \cite{PecoraEA26PUNCH} is not directly equivalent to the transverse correlation length considered here, and a direct comparison between the two is not straightforward.
Meanwhile, we note that the correlation scales examined here are 
substantially 
larger than those reported with in situ measurements,
an effect likely influenced by 
spatial resolution of PUNCH instrumentation. 
The correlation scale in the annulus $70$--$90\,R_\odot$ is about $1.5R_\odot (\approx1.0\times 10^6\rm{km})$, while the correlation scales from Parker Solar Probe measurements at $0.4\rm{AU}$ \citep{CuestaEA22-iso} is about $1.0\times 10^5 \rm{km}$. 

We cannot fully explain these contrasts, but they may potentially arise from differences in the measurements themselves. First, most previous studies using in-situ measurements investigated the correlation function of the magnetic field, while we are looking at the brightness, which is related to electron density in a complicated way (more details in the following section). Second, the PUNCH images and the in-situ measurements focus on different ranges of scales. The PUNCH images capture the ultra-large fluctuations at large scales (the correlation scale and the upper end of the inertial range) \cite{pecora2024magnetohydrodynamic}, while the in-situ measurements move toward the more strongly-coupled nonlinear cascade that begins at the correlation scale. Third, the studies using in-situ measurements measure variations parallel and perpendicular to the mean magnetic field, while here we assign the radial and transverse directions as parallel and perpendicular in the plane of the sky. Although for in-situ measurements closest to the Sun, the magnetic field is dominantly radial, the Parker-spiral average magnetic field \citep{Parker58a} changes direction with varying radial distance \citep{CuestaEA22-iso} and the radial direction in this case means directions into and out of the plane of sky. Fourth, in-situ estimates of the correlation scale rely on Taylor's frozen-in approximation \citep{Taylor38} to convert single-point temporal lags into spatial lags; in contrast, we directly sample multi-point spatial data. Fifth, the trend of correlation anisotropy with heliocentric distance could also be affected by decreasing signal-to-noise ratio with increasing helioradius. Random noise would decorrelate faster than the real signal.

Several technical aspects of the present analysis may influence the quantitative correlation lengths and their inferred radial evolution, as well. Negative-valued pixels are excluded from the analysis, and their spatial distribution
may vary across the image, which may cause additional uncertainty in the image. The removal of the slowly varying background by Gaussian smoothing introduces an additional characteristic spatial scale, so the inferred correlation lengths may also depend to some degree on the adopted width. In addition, within each annulus the number of valid pixel pairs decreases toward larger spatial lags because of the finite annular geometry and the presence of masked pixels. The long-lag part of the correlation function is therefore generally less well sampled and may be more susceptible to statistical fluctuations. 
Finally, several analysis parameters, including the annular widths, lag range and spacing, masking criteria, normalization procedure, and the criterion used to determine the correlation length, are empirically chosen and may introduce additional systematic uncertainty.

We cannot rule out the possible ambiguities in the calculation of correlation function. For turbulence analyses, an interval size containing several correlation scales is ideal. Further, the interval should not be so large that nonstationarity and inhomogeneity such as solar rotation effects are included \citep{wang2026interplanetary}. \citet{IsaacsEA15,JagarlamudiApJ19} showed that estimates of the autocorrelation function of magnetic fluctuations in the solar wind exhibit a strong dependence on the choice of interval length, and the optimized averaging window for computing correlation scales was determined to be between 10 and 20 $\rm{hr}$ long at  $1\rm{AU}$. More recently, \cite{perez2026integral} proposed a new ergodicity-based methodology to unambiguously estimate the correlation scale, which could refine previous work using in-situ observations and our analyses here. However, a fundamental assumption in most analyses of spacecraft observations,  which is almost impossible to avoid when comparing with turbulence theories, is stationarity (for in-situ measurements) or homogeneity (for PUNCH images we analyze here). 

Finally, there are unavoidable complications pertaining to the observed images. 
First, the signal incorporates information that is 
composed of Thomson-scattered light from a range of line-of-sight depths centered about the Thomson sphere \citep{vourlidas2006proper,deforest2013thomson}.
Ongoing analysis by the PUNCH team and others 
is expected to make progress in better understanding 
this ``line-of-sight averaging'' effect.
However, for these first results, the analysis of the  
original data does not take into account any possible refinements.
A companion paper \citep{PecoraEA26PUNCH}
takes into account some line-of-sight refinements for 
spectral shapes in an empirical approach based on forward modeling of simulation data \citep{pecora2024magnetohydrodynamic}.
However, such approximate corrections are not expected to influence the observed development of directional anisotropies, which has been the present emphasis. 
Second, the observed images
owe their 
direct cause to 
Thomson scattering from electron density inhomogeneities. 
Although Thomson-scattered white light images could also be interpreted as depictions of 
magnetic field lines and 
flux structures, such an interpretation should be made with appropriate caveats.
This interpretation relies to a great extent on the plasma motions being constrained by magnetic forces and by the magnetic topology, so that plasma can flow along flux tubes rather easily. However, plasma motions across field lines (transverse flows) can occur, particularly at high plasma $\beta$ (the ratio of thermal pressure to magnetic pressure), which distort frozen-in flux.
PUNCH images span regions in
which plasma $\beta$  presumably 
varies to a substantial degree, from low in the sub-Alfv\'enic corona to instances of $\beta$ potentially of the order of, or greater than, unity in certain regions \citep{Gonzalez2026ApJ}. 
In this regard, it is 
best to bear in mind that the observed correlations are those of electron density, which have been observed to behave distinctly from magnetic correlations in a recent in-situ study \citep{WangEA24-density}; further inferences regarding plasma turbulence, magnetic structure and classical hydrodynamic or MHD turbulence 
may require more caution.

Even in light of these complicating issues
the PUNCH dataset affords an indispensable opportunity to break essentially new ground in the exploration of
correlation anisotropy in heliospheric plasmas. 
A straightforward extension of our work is to compare 
polarized and unpolarized images, which we leave for future work. Because the scattering affects the polarized brightness distinctly from the total brightness, this provides an additional constraint to test our results against in future work.

\begin{acknowledgments}

This research is supported in part by NASA's PUNCH mission under subcontract
N99054DS to the University of Delaware and by a Plan for NASA EPSCoR Research Infrastructure Development (RID) in Delaware (NASA award 80NSSC22M0039). S.E.G. acknowledges the support of the National Center for Atmospheric Research, a major facility sponsored by the U.S. National Science Foundation under Cooperative Agreement No. 1852977.

\end{acknowledgments}


\bibliographystyle{aasjournalv7}
\bibliography{arxiv_refs}



\end{CJK*}
\end{document}